\documentclass[superscriptaddress,twocolumn,
 amsmath,amssymb,
 aps,
]{revtex4-1}
\usepackage{import}
\usepackage{braket}
\usepackage{graphicx}
\usepackage{dcolumn}
\usepackage{siunitx}
\usepackage{bm}
\usepackage{color}
\usepackage{pgfplots}
\usepackage{hyperref}

\definecolor{d_Red}{RGB}{190, 30, 45}
\definecolor{d_Cyan}{RGB}{0, 174, 239}
\definecolor{d_Gold}{RGB}{175, 169, 97}
\definecolor{d_Yellow}{RGB}{255, 213, 58 }
\definecolor{d_Purple}{RGB}{104, 36, 109 }
\definecolor{d_Heather}{RGB}{203,168,177 }
\definecolor{d_Stone}{RGB}{218,205,162}
\definecolor{d_Sky}{RGB}{165,200,208}
\definecolor{d_Cedar}{RGB}{182,170,167}
\definecolor{d_Concrete}{RGB}{179,189,177 }
\definecolor{d_Ink}{RGB}{0,42,65 }
\definecolor{d_Black}{RGB}{51,49,50 }

\usepackage{mathtools}
\newcommand{\JQCAffiliation}{\affiliation{
Joint Quantum Centre (Durham-Newcastle), Department of Physics, Durham University, South Road, Durham, DH1 3LE, United Kingdom
}}

\newcommand{\SKLQOAffiliation}{\affiliation{
 State Key Laboratory of Quantum Optics and Quantum Optics Devices, Institute of Laser Spectroscopy, Shanxi University, Taiyuan 030006, China
}}

\begin{document}

\preprint{APS/123-QED}

\title{Single-Photon Stored-Light Interferometry}

\author{Yuechun Jiao}
\JQCAffiliation
\SKLQOAffiliation

\author{Nicholas L. R. Spong}
\JQCAffiliation

\author{Oliver D. W. Hughes}
\JQCAffiliation 

\author{Chloe So}
\JQCAffiliation

\author{Teodora Ilieva}
\JQCAffiliation

\author{Kevin J. Weatherill}
\JQCAffiliation 

\author{Charles S. Adams}
\JQCAffiliation

\date{\today}

\begin{abstract}

We demonstrate a single-photon stored-light interferometer, where a photon is stored in a laser-cooled atomic ensemble in the form of a Rydberg polariton with a spatial extent of $10 \times1\times1\si{\micro\metre}^3$. The photon is subject to a Ramsey sequence, i.e. `split' into a superposition of two paths. After a delay of up to 450~ns, the two paths are recombined to give an output dependent on their relative phase. The superposition time of 450~ns is equivalent to a free-space propagation distance of 135~m. We show that the interferometer fringes are sensitive to external fields, and suggest that stored-light interferometry could be useful for localized sensing applications.

\end{abstract}

\maketitle

\section{Introduction}

 \maketitle


Interferometers \cite{Opticsf2f} are a versatile tool in engineering, oceanography, seismology, metrology and astronomy \cite{Abbott2016}.
Replacing photons by massive particles creates an interferometer  \cite{Adams1994} that is sensitive to gravity and other inertial effects.
Interferometry using the internal quantum states of atoms \cite{Ramsey1950,Ramsey1990} has become an important technique in the measurement of time and quantum coherence. Hybrid light-matter interferometers are also possible. In a slow-light interferometer \cite{Shi2007}, the field is partly photonic and partly atomic. A hybrid atom-light interferometer using internal atomic states to form a beam splitter has also been demonstrated \cite{Shuying2015}.  Interferometry using atomic Rydberg states is useful in the detection of electric fields, as demonstrated using individual atoms \cite{Facon2016}, atomic beams \cite{Palmer2019} and Rydberg-dressed cold-atom ensembles \cite{Arias2019}. For all types of interferometry, the measurement sensitivity is limited by the time between the splitting and recombination processes, which is typically proportional to the length of the interferometer arms.

Here, we propose and demonstrate a hybrid interferometer based on a single photon stored as a spin wave in a cold atomic gas. The advantage of using stored light (or slow light) is that superposition time is decoupled from the size of the interferometer. For example, we demonstrate a quantum superposition time of 450~ns---equivalent to a free-space interferometer path length of 135 metres---using an inferterometer localized to a few microns. The two paths of the interferometer are formed by driving the spin wave into a superposition of quantum states. We employ highly-excited Rydberg states which allows the splitting and recombining of the optical paths to be performed using a microwave field. Similar to other Rydberg experiments  \cite{Facon2016,Palmer2019,Arias2019}, the output is sensitive to DC and AC electric fields. As the interferometer is localized over a length scale of only a few microns, the Rydberg blockade mechanism \cite{Lukin2001} forces the interferometer to operate with only one photon at a time. This has the advantage that phase shifts due to atom-atom interactions are suppressed. Both localisation and the single-photon character may be advantageous in some measurement applications, such as sensing and imaging  \cite{Sedlacek2012, Fan2015, Facon2016, Wade2017}.

\section{Experimental Demonstration}

\begin{figure}[t!]
\centering
\includegraphics[width=\linewidth]{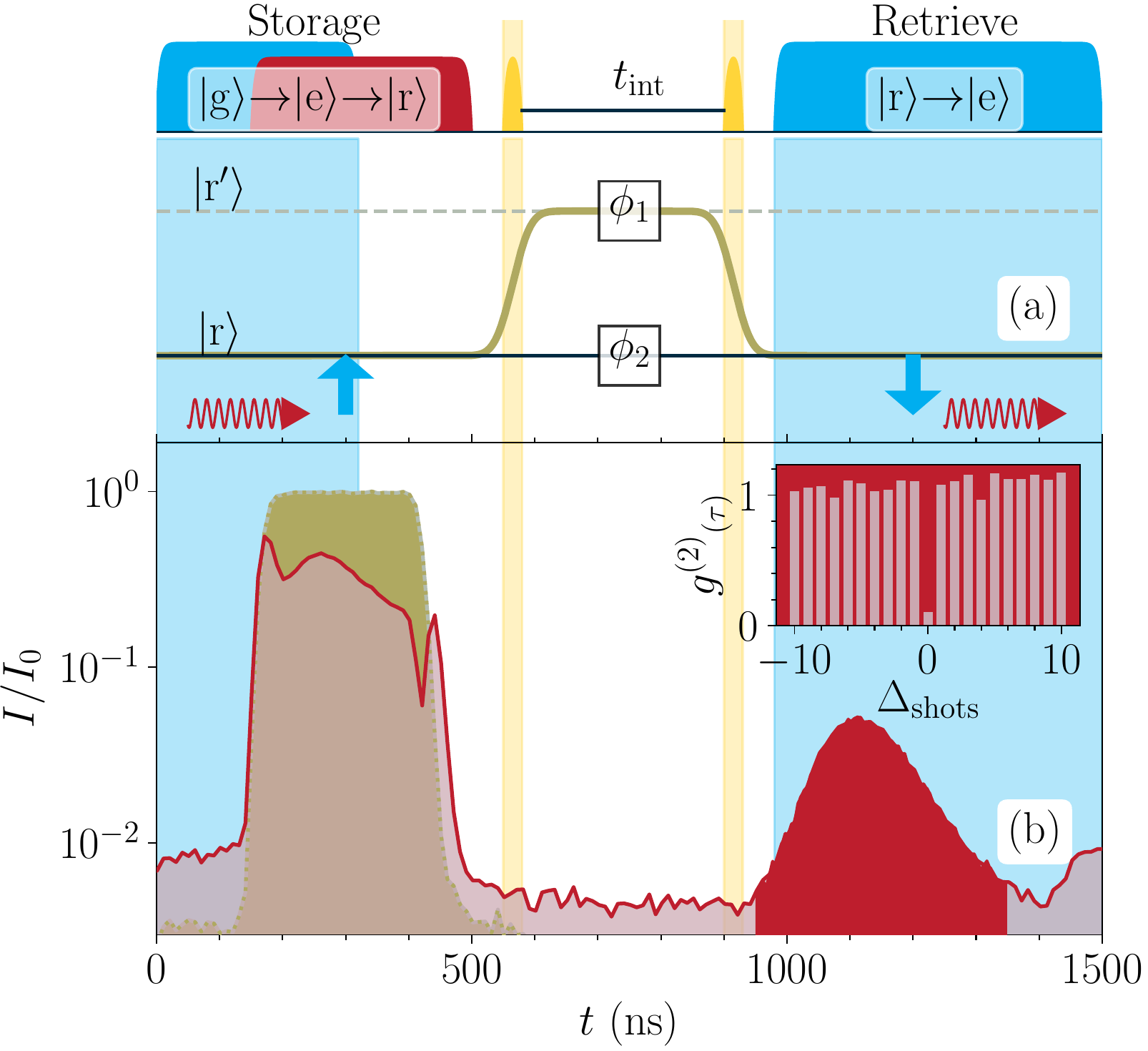}
\caption{ {\bf Principle of the interferometer}: (a) A photon (red arrow)---incident on a cold atom ensemble, and resonant with the $\vert{\rm g}\rangle\rightarrow\vert{\rm e}\rangle$ transition---is stored as an excitation in the  Rydberg state  $\vert{\rm r}\rangle$, using a coupling field (blue arrow) resonant with the $\vert{\rm e}\rangle\rightarrow\vert{\rm r}\rangle$ transition. A microwave pulse (yellow) creates a superposition of two paths corresponding to internal quantum states, $\vert{\rm r}\rangle$ and $\vert{\rm r'}\rangle$. The superposition is maintained for  a free evolution time, $t_{\rm int}$, and then recombined via a second microwave pulse. Finally, the population in $\vert{\rm r}\rangle$ is read out by  driving the  transition $\vert{\rm r}\rangle\rightarrow\vert{\rm e}\rangle$ (blue). (b) Intensity measured during the complete sequence: storage ($t=120-420$~ns), interferometer ($t=560-920$~ns) and read-out $t=950-1350$~ns. Inset: Coincidence counts in a single shot and between shots indicating the single photon character of the stored-light. For a Rydberg state $\vert{\rm r}\rangle$ with principal quantum number $n=90$.
}
\label{fig:fig1}
\end{figure}

The principle of our single-photon stored-light interferometer is illustrated in Fig.~\ref{fig:fig1}. To store a photon, first laser-cooled $^{87}{\rm Rb}$ atoms are loaded into an 862~nm optical dipole trap with a waist $w_r= 4.5\,\si{\micro\metre}$ and a trap depth of $\sim 0.5~{\rm mK}$. The atoms are optically pumped into the $\ket{\rm g}=\ket{5\textrm{S}_{1/2}, F = 2, m_F = 2}$ sublevel.
Next, we illuminate the laser-cooled atomic ensemble with both a probe laser (780~nm) and a control laser (480~nm) for 500~ns.  The probe beam, resonant with $\ket{\rm g}=\ket{5\textrm{S}_{1/2}, F = 2, m_F = 2} \rightarrow \ket{\rm e} = \ket{\textrm{P}_{3/2}, F' = 3, m'_F = 3}$ transition is focused into the ensemble with a beam waist of \SI{1.5}{\micro\metre}. The control beam, resonant with the $\ket{\rm e}\rightarrow\ket{\rm r}=\ket{n \textrm{S}_{1/2}}$ transition is focused to a waist of \SI{25}{\micro\metre}. The probe and control beams are circularly polarized and counter-propagate. Before the probe is switched off, the control is ramped down to zero \cite{Moehl2020}. The spatial extent of the stored photon is of order $10\times1\times1\,\si{\micro\metre}^3$. 

After storage, the stored photon is `split' using a microwave field that couples the highly-excited Rydberg state, $\ket{\rm r}=\ket{n \textrm{S}_{1/2}}$, to  another Rydberg state $\ket{\rm r'} = \ket{n\textrm{P}_{3/2}}$, shown in yellow in Fig.~\ref{fig:fig1}(a). The microwave field amplitude is uniform over the dimensions of the stored photon.   The microwave pulse, lasting 25~ns, is calibrated as a $\pi$/2 pulse to realise the superposition state $\frac{1}{\sqrt{2}}(\vert{\rm r}\rangle+\vert{\rm r'}\rangle)$.  After a period of free evolution, $t_{\rm int}$, the microwave source drives another $\pi$/2 pulse to recombine the two paths.  Finally, the population in $\vert{\rm r}\rangle$ is measured by coupling this state back to $\vert{\rm e}\rangle$ using the control field (shown in blue in Fig.~\ref{fig:fig1}). Each experimental run is performed thousands of times. A complete run consists of laser cooling and trapping for 120~ms followed by 20,000 individual interferometer measurements performed on the same ensemble in 90~ms. This gives an effective repetition rate of 95~kHz. All the cooling and trapping lasers are turned off during the interferometer measurements. The transmission of probe beam is measured throughout using a single-photon counter,  see  Fig. \ref{fig:fig1}(b). More details on experimental apparatus and the relevant atomic levels can be found in previous work \cite{Busche2016,Busche2017,Moehl2020}. 

We employ highly-excited Rydberg states with principle quantum numbers in the range $n=60-90$. The Rydberg blockade mechanism \cite{Lukin2001,Li2016,Ornelas-Huerta2020} limits the inteferometer to only one photon at a time. To demonstrate this we perform a Hanbury Brown Twiss (HBT) measurement on the light retrieved from the ensemble. The retrieved light is  split by a 50:50 beam splitter and sent to two detectors. The normalised coincidence counts for a single retrieval is strongly suppressed, see Fig. \ref{fig:fig1}b(inset). For a Rydberg state $|{\rm r}\rangle$ with principal quantum number $n=90$, the probability to observe two photons in the same experimental shot, characterised by the normalized second-order intensity correlation, is 
$g^{(2)}(0) = 0.10\pm0.02$.  
The correlation function between successive shots $g^{(2)}(\tau)$ is unity as expected.

\begin{figure}[htbp]
\centering
\includegraphics[width=\linewidth]{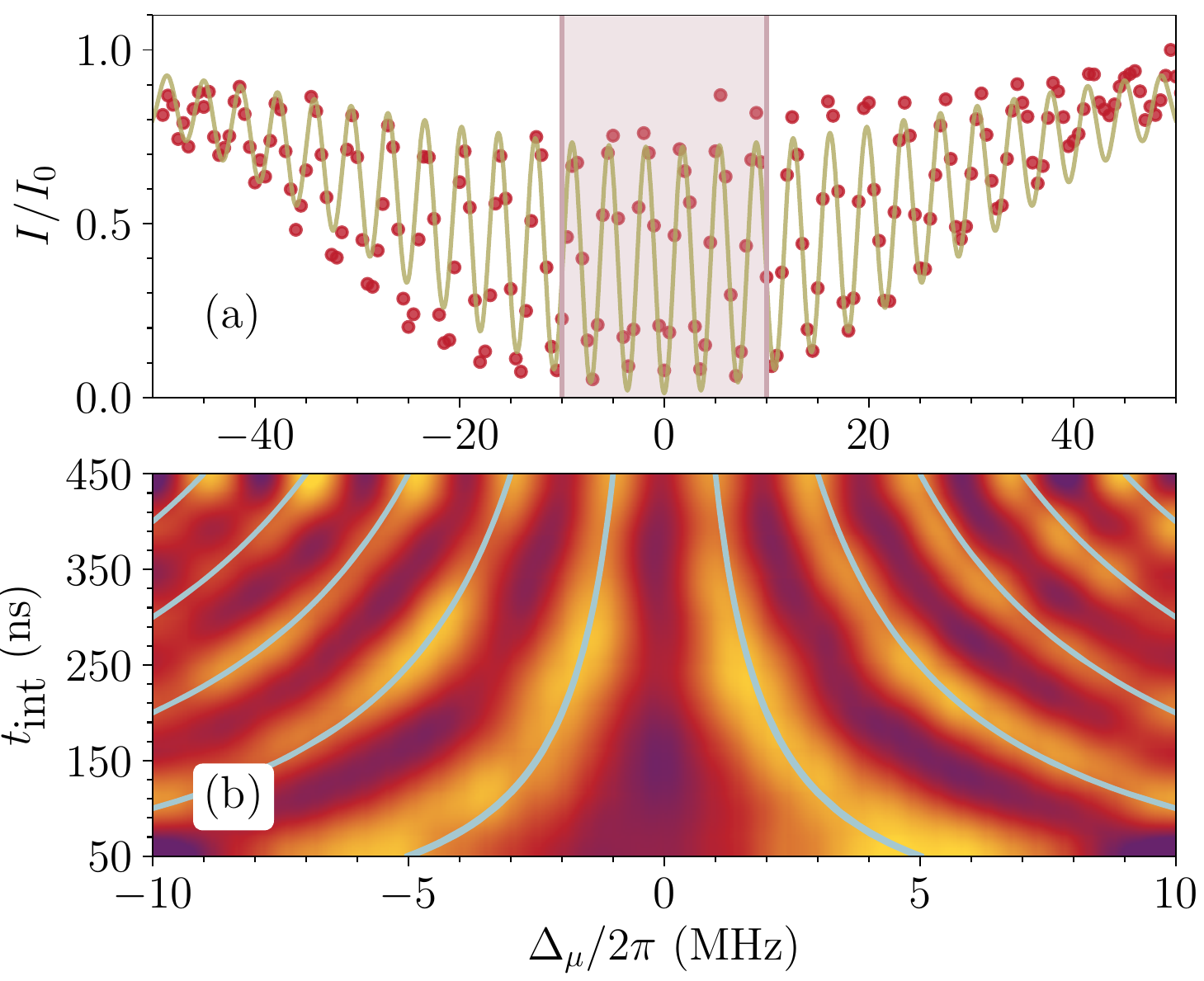}
\caption{\textbf{Operation of the interferometer}: (a) The output fringes of the single-photon interferometer by varying the detuning of the microwave field $\Delta_\mu$. The data (red dots) is obtained for a superposition time $t_{\rm int}=250$~ns. The data is fitted by solving the optical Bloch equations for the complete time sequence. Apart from an overall envelope the fit has the form of Eq.~\ref{eqn:int}. (b) Measured interferometric fringes as a function of both $\Delta_\mu$ and the superposition time $t_{\rm int}$, for $t_{\rm int}\leq 450$~ns. For this data we use $\ket{\rm r} = \ket{60 \textrm{S}_ {1/2}}, \ket{\rm r'}  = \ket{59P_{3/2}}$. The lines show the theoretically  predicted retrieval maxima.}
\label{fig:fig2}
\end{figure}

The output of the interferometer has the form \cite{Opticsf2f}
\begin{equation}
{\cal I}={\cal I}_0(1-\cos\phi)~,
\label{eqn:int}
\end{equation}
where $\phi=\Delta_\mu t_{\rm int}$ and $\Delta_\mu=[\omega_\mu-(E_{\vert {\rm r}'\rangle}-E_{\vert {\rm r}\rangle})/\hbar]$ is the detuning of the microwave field. Consequently, to measure the interference fringes we can either vary the microwave field frequency, $\omega_\mu$, or the superposition time, $t_{\rm int}$.  Note the phase $\phi$ and hence the intensity output are also sensitive to any perturbation in the energy levels, $E_{\vert {\rm r}\rangle}$ and $E_{\vert {\rm r}'\rangle}$.  The interference fringes as a function of both the microwave field detuning, $\Delta_\mu$, and the superposition time, $t_{\rm int}$, are shown in Fig. \ref{fig:fig2}. In Fig. \ref{fig:fig2}(a) we show the fringes for $t_{\rm int}=250$~ns.
In the experiment, the visibility of the interference fringes is reduced by technical factors including the imperfect removal of population from the Rydberg manifold after each shot of the experiment, and imperfect polarization of the microwave fields. Future work will focus on improving addressing these technical issues. In Fig. \ref{fig:fig2}(b) we focus on  the interference fringes in the range $\Delta_\mu/(2\pi)=-10$ to 10~MHz, and show the effect of varying the superposition time, $t_{\rm int}$. The data are overlaid with theoretical retrieval maxima in yellow. As expected, as we increase $t_{\rm int}$, the interferometer becomes more sensitive to changes in the relative phase. In the current experiment, $t_{\rm int }$ is limited to of order 1000~ns by motional dephasing \cite{Busche2017}, which scrambles the information about the stored photon.

\begin{figure}[h!]
    \centering
    \includegraphics[width=\linewidth]{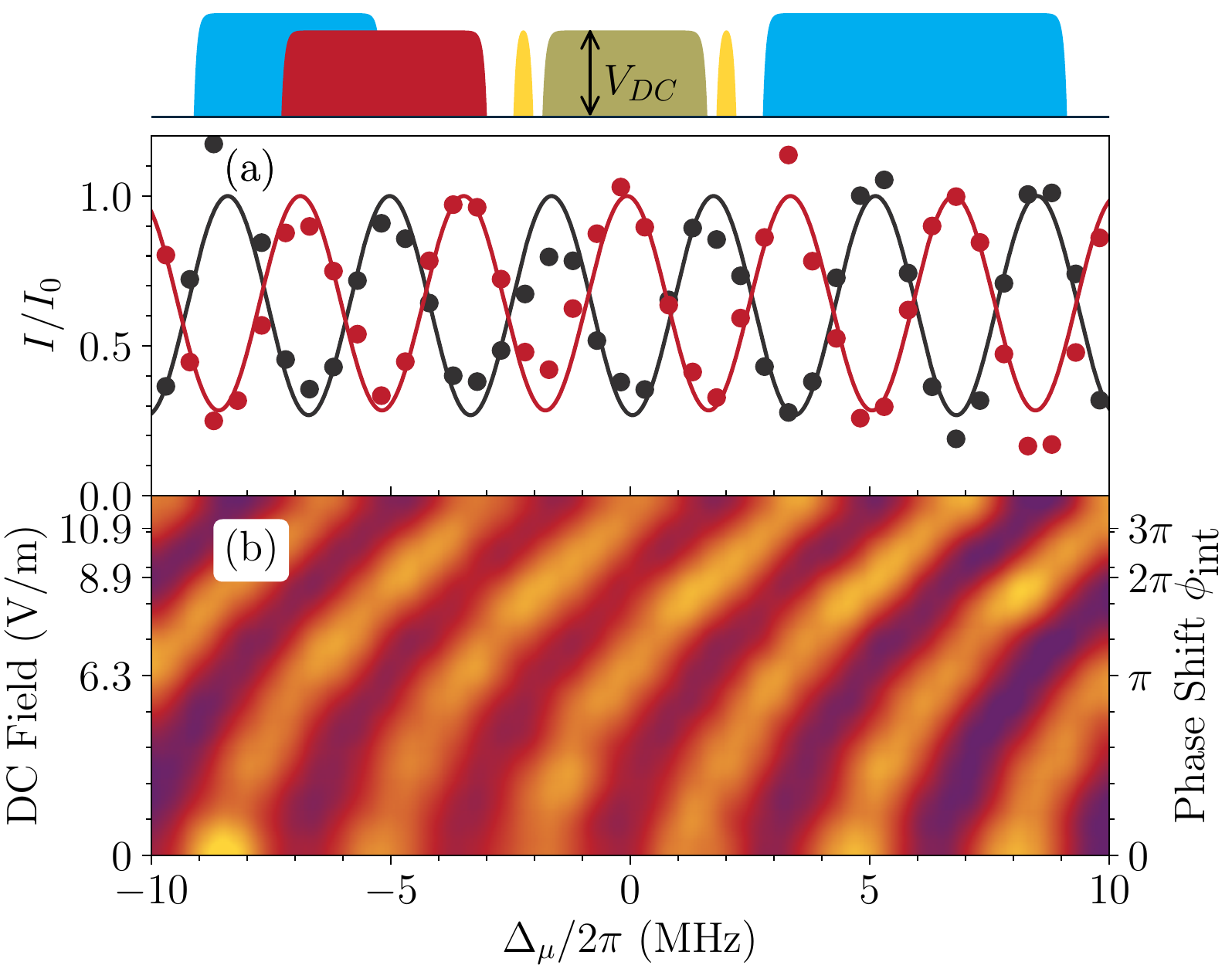}
    \caption{\textbf{Sensitivity of the single-photon interferometer to a DC electric field}. Top: The timing sequence is the same as in Fig.~1, i.e. photon storage, interferometer (yellow pulses) and retrieval, but with the addition of DC electric field during the superposition time of the interferometer (green).  (a) Measured interference fringes at a DC field of 0~V (black dots) and 6.3~V/m (red dots). The data are normalized to the zero-field interference fringe maxima. The solid line is a fit. (b) Colormap showing the electric field dependent phase shift of the interferometer fringes. The value of DC fields is calibrated using the known polarizabilities of the Rydberg states \cite{ARC}.}
    \label{fig:fig3}
\end{figure}

Next, we demonstrate the sensitivity of the interferometer fringes to external fields. First, we apply a DC electric field. This has the effect of shifting the energy levels due to the DC Stark effect, i.e.
\begin{equation}
\Delta E_{\vert{\rm r}\rangle} = -\textstyle{\frac{1}{2}} \alpha_{\vert{\rm r}\rangle}{\cal E}^2~,
\label{eqn:DC}
\end{equation}
where $\alpha_{\vert{\rm r}\rangle}$ is the polarizability of the state $\vert{\rm r}\rangle$, and ${\cal E}$ is the applied electric field.
The external field is applied during the superposition time between the two microwave $\pi/2$-pulses, see Fig. \ref{fig:fig3}(top). Fig.~\ref{fig:fig3}(a) shows the interference fringes at zero field and a DC field of 6.3~V/m. At this field value the phase shift is $\pi$.  Note that the fringe contrast is conserved, as the DC field only induces a global phase shift, without perturbing the spatial mode of the photon. Figure~\ref{fig:fig3}(b) shows the fringe shift from 0 to 12~V/m, with the phase shift indicated on the right-hand axis. As expected we observe a quadratic shift of the Ramsey fringes matching the relative dc-Stark shift of states $\vert{\rm r}\rangle$ and $\vert{\rm r'}\rangle$. 

Second, we investigate the sensitivity of our single-photon stored-light interferometer to an external AC electric field. In the case of AC fields, the polarizabilities, $\alpha_{\vert {\rm r}\rangle}$ and $\alpha_{\vert {\rm r}'\rangle}$, exhibit resonances at particular rf, microwave and terahertz frequencies that match electric dipole allowed transitions between atomic Rydberg states, and these resonances have emerged as a promising platform  for sensitive measurements of rf \cite{Jing2020} and terahertz \cite{Downes2020} fields. 
The sensitivity of the interferometer to a near-resonant external microwave field is predicted by the AC stark shift of individual Rydberg states. The energy shift of a state $\vert i\rangle$ can be represented by
\begin{equation}\Delta E_i = - \frac{\mathcal{E}^2}{4\hbar}\sum_{i \neq f}\vert\langle i\vert\mathbf{d}\vert f\rangle\vert^2    \left( \frac{1}{\omega_{\mu 2}-\omega_{if}}+\frac{1}{\omega_{\mu 2}+\omega_{if}}\right)~,\label{eqn:AC}\end{equation}
where $\langle i\vert\mathbf{d}\vert f\rangle$ is the electric dipole moment for a transition between states $\vert i\rangle$ and $\vert f\rangle$, $\omega_{\mu 2}$ is the frequency of the external field, and $\omega_{if}$ is the resonant transition frequency between $i$ and $f$.
In the experiment, we use an interferometer based on the states $\vert{\rm r}\rangle = 60 S_ {1/2}$ and $ \vert{\rm r}'\rangle = 59P_{3/2}$. The external microwave field is chosen to couple $\vert i\rangle=\vert{\rm r}'\rangle  = 59P_{3/2}$ to another Rydberg state $\vert f\rangle=\vert{\rm r}''\rangle  = 59S_{1/2}$. The resonant frequency for this case is 18.2~GHz and the transition has a dipole moment, $\langle i\vert\mathbf{d}\vert f\rangle = 4107$~Debye \cite{ARC}. The pulse sequence is shown in Fig.~4(top). The additional microwave field (green) is applied during the interferometer superposition time between the two $\pi/2$-pulses (yellow). Figure~4{\bf a} shows measurements of the single-photon Ramsey fringes as a function of both the detuning of the additional microwave field, $\Delta_{\mu2}=\omega_{\mu 2}-\omega_{if}$, and the superposition time, $t_{\rm int}$. Fig.~4(b) shows how the shift of the interference fringes follows a dispersive-like law as expected. A fit to equation equation \ref{eqn:AC}---the dotted line in the figure---is in good agreement with the data.

An interesting feature of the stored-light interferometer is that it provides two ways of measuring an external rf field, first via a fringe shift as shown in Fig.~4, and second via the loss of visibility. Figure~5 shows rf sensing data in the vicinity of a resonance. Figure~5(a) show the Ramsey fringes as function of the detuning of the microwave field, $\Delta_{\mu}$, while varying detuning of the external rf field, $\Delta_{\mu2}=\omega_{\mu 2}-\omega_{if}$. In contrast, to the DC field case, Fig.~3, now we observe a change in the fringes visibility in the vicinity of the microwave resonance. When the external field is close to a resonance with another Rydberg state, we see a dramatic loss of fringes contrast. This is plotted in Fig.~5{\bf b}. The loss occurs because there is a probability that a part of the photon energy is coupled into microwave field similar to the classic decoherence experiment of Brune {\it et al}. \cite{Brune1996}. The data is fitted with a lineshape that is consistent with Fourier transform of the microwave pulse duration of 200~ns. Compared to Fig.~4, this plot illustrates how the loss of coherence is more frequency selective than the phase shift, and could be used for high-resolution measurements of Rydberg transition frequencies, which are important in the determination of quantum defects \cite{Li2003}.

\begin{figure}[htbp]
    \centering
    \includegraphics[width=\linewidth]{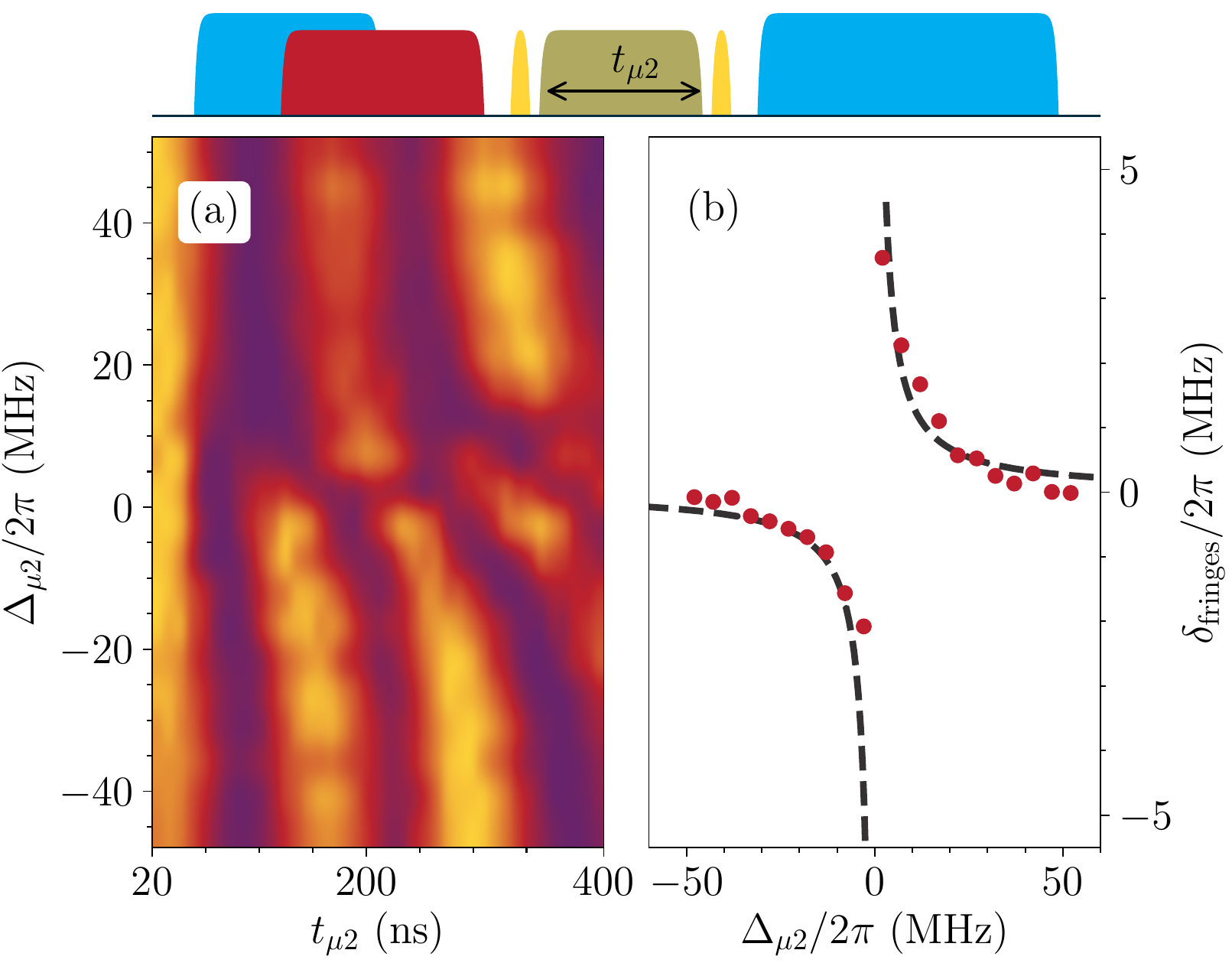}
    \caption{{\bf Single-photon sensing of an external microwave field:} Top: Timing sequence as previously except now with an additional microwave field applied for a time $t_{\mu 2}$ during the interferometer superposition time.
    (a) Colormap showing measurements of the single-photon interference fringes as a function of both the detuning of the external field, $\Delta_{\mu2}$, and the inteferometer superpositon time, $t_{\rm int}$ with $\Delta_\mu/(2\pi)=8$~MHz. The external microwave field couples one path of the interferometer to another Rydberg state.  (b) Shows that the frequency shift of the Ramsey fringes, $\delta_{\rm fringes}$, follows a dispersive-like law as expected with detuning of $\Delta_{\mu2}$. The dotted line shows a fit to equation \ref{eqn:AC}.}
    \label{fig:fig4}
\end{figure}

Applications of the single-photon interferometer will be explored in further work. For sensing applications, the current experiment is limited by shot noise (in contrast to classical measurements \cite{Arias2019}, quantum measurement is limited to only one photon per shot) and photon storage time. The latter could be greatly increased by eliminating motional dephasing of the spin wave, either by using Doppler-free configuration \cite{Sibalic2016}, or using an optical lattice to localise the atoms \cite{Schnorrberger2009}. Ultimately, the fundamental limit is the lifetime of the Rydberg states which for circular states can be $>1000$~s \cite{Nguyen2018}. Using these techniques, a stored-light interferometer could rival state-of-the-art electrometers \cite{Facon2016,Arias2019,Jing2020}.

\section{Conclusion}

In summary, we have realized a single-photon stored-light Ramsey-type interferometer via storing optical photons as Rydberg polaritons. The character of the Ramsey-like interference fringes has been demonstrated. We show that the interferometer, localized to a length scale of just a few microns, can be used to measure external fields. In these examples, information about the field is mapped onto the quantum state of the stored photon. This can lead to both changes in the fringe position and visibility. 
Our method paves the way towards advanced quantum probing of external fields. Potential applications include precision measurement of Rydberg transition frequencies, and sensing any field that perturbs either of the Rydberg states involved. The interferometer can be made sensitive to single optical photons by mapping them into Rydberg polaritons \cite{Paredes2014}, and could be used as the basis to realise a photonic phase gate \cite{Tiarks2019}.

    \bigskip
    
    \noindent Funding and Acknowledgements. We acknowledge support from Engineering and Physical Sciences Research Council (EPSRC) Grants EP/M014398/1, EP/R002061/1 and EP/S015973/1. The figure data are available on the Durham University Collections repository (\href{http://dx.doi.org/10.15128/}{doi:10.15128/r2r207tp335}).

    \bigskip

\begin{figure}[h]
    \centering
    \includegraphics[width=\linewidth]{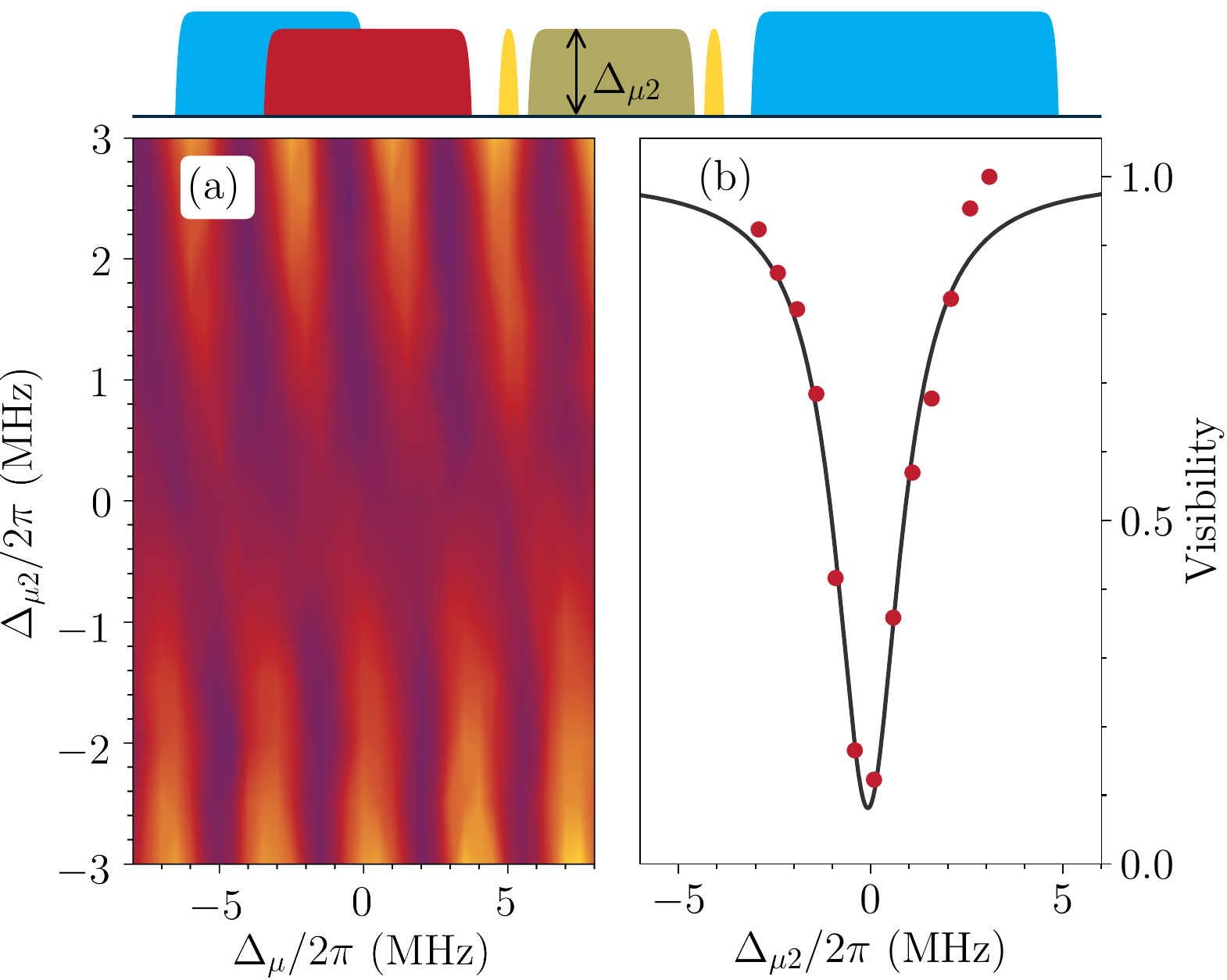}
    \caption{{\bf Effect of an external rf field on the visibility of the single-photon interference fringes:} Top: Timing sequence same as Fig.~4 except that the duration of the external field is now fixed at $t_{\mu 2}=200$~ns. (a) Colormap showing the single-photon interference fringes as a function $\Delta_{\mu}$ and the detuning of the external field, $\Delta_{\mu2}$,  with $t_{int}=250$~ns. On this scale, $\Delta_{\mu2}=-3$ to $3$~MHz, the AC Stark shift seen in Fig.~4 is scarcely visible, whereas the loss of coherence is pronounced.
    (b) Plot of the fringe visibility in the vicinity of resonance with the external field. The data is fitted with a lineshape consistent with Fourier limited linewidth for a 200~ns rf pulse.}
    \label{fig:fig5}
\end{figure}

\bibliography{apssamp}

\end{document}